\documentclass[%
amsmath,amssymb,%showkeys,
aps,
prl,
twocolumn,
]{revtex4-2}

\usepackage{graphicx}% Include figure files
\usepackage{dcolumn}% Align table columns on decimal point
\usepackage{bm}% bold math
\begin{document}
%\preprint{APS/123-QED}

\title{Mass-imbalanced SU(N) Fermi gases}
\author{Jordi Pera}
\email{jordi.pera@upc.edu}
\author{Joaquim Casulleras}
\email{joaquim.casulleras@upc.edu}
\author{Jordi Boronat}
\email{jordi.boronat@upc.edu}
\affiliation{%
 Departament de F\'\i sica, Campus Nord B4-B5, Universitat Polit\`ecnica de 
Catalunya, E-08034 Barcelona, Spain
}%

\date{\today}% It is always \today, today,

\begin{abstract}
We report a fully analytical description of zero-temperature itinerant ferromagnetism in repulsive SU(N) Fermi gases with arbitrary mass imbalance among components. Using  
perturbation theory in the gas parameter $x=k_F a_0$, with $k_F$ the Fermi 
momentum and $a_0$ the s-wave scattering length, we derive the  
second-order energy for arbitrary spin polarization and arbitrary mass ratio. 
Our main result is a closed analytic expression for the  
beyond-mean-field correction in mixtures with unequal masses. 
This analytical result extends the theory of dilute Fermi gases beyond the 
mass-balanced case and provides a compact equation of state for multicomponent 
mixtures. We show that mass imbalance breaks the paramagnetic symmetry already 
in the non-interacting limit, favors the occupation of heavier components, and lowers the interaction strength required to reach a fully polarized state. For $S=1/2$, 
the system evolves continuously from a mass-induced partially polarized state to 
full ferromagnetism. For larger spins, distinct mass distributions generate 
qualitatively different polarization paths, including smooth and discontinuous 
sequences. Our results identify mass imbalance as a powerful control tool for 
magnetic ordering in ultracold Fermi mixtures.
\end{abstract}

\maketitle

%\tableofcontents

%\section{Introduction}
Itinerant ferromagnetism in many-body repulsive Fermi systems is a paradigmatic 
quantum phase transition in which a paramagnetic state becomes magnetically 
ordered as interactions increase~\cite{Stoner1933}. In the simplest Stoner 
picture, 
this transition follows from the competition between kinetic energy, which 
favors equal spin populations, and interaction energy, which favors population spin 
segregation. As in other quantum phenomena, ultracold atomic gases provide a 
particularly clean platform to investigate this phase transition because 
interactions, dimensionality, and spin degeneracy can be tuned experimentally 
with high precision~\cite{Giorgini2008}. However, observing the repulsive 
ferromagnetic branch remains challenging due to its metastability, that leads 
to non-Fermi liquids~\cite{pfleiderer}, crystallization~\cite{leduc} or 
formation of dimers~\cite{Sanner2012} before reaching the ferromagnetic state. 
Recently, evidence of ferromagnetic behavior has been reported through 
spin-domain inmiscibility in ultracold Fermi gases~\cite{Valtolina2017}.

A promising route toward richer magnetic behavior is to consider multicomponent 
Fermi gases. 
Such systems are nowadays experimentally accessible in alkaline-earth-like 
atoms, such as $^{173}$Yb with $S=5/2$~\cite{Pagano2014} and $^{87}$Sr with 
$S=9/2$~\cite{Goban2018}, that realize SU(N) 
symmetry. Previous theoretical work has shown that increasing the spin degeneracy of the gas lowers the critical density for the ferromagnetic transition~\cite{Pera1,Pera2,Pera3}. 
Another path is mass imbalance, which 
explicitly breaks spin symmetry and modifies the competition between kinetic and 
interaction energies, promoting polarization mechanisms distinct from those in mass-balanced gases~\cite{Cui2013}, and reducing the rate of three-body 
recombinations~\cite{Petrov2003}. Experimentally, at least two examples of 
Fermi-Fermi mixtures with mass imbalance have been produced: 
$^6$Li-$^{40}$K~\cite{Kohstall2012}, $^6$Li-$^{133}$Cs~\cite{Lippi2024}, and 
$^{161}$Dy-$^{40}$K~\cite{Ravensbergen2020}.

Previous theoretical work has shown that unequal masses can enhance phase 
separation and magnetic 
ordering in repulsive Fermi mixtures~\cite{Cui2013,Fratini2014}. Related 
studies in nuclear and multicomponent Fermi systems have also emphasized the 
role of mass imbalance in modifying correlation effects and the equation of 
state~\cite{Pethick2025}. However, a fully analytical second-order equation of 
state for arbitrary spin polarization and arbitrary mass ratio has remained 
elusive. This missing ingredient is crucial because the dependence of the 
energy on both polarization and mass ratio determines the actual magnetic 
configuration of the system.

In this Letter, we derive the  energy of a repulsive 
SU(N) Fermi gas, for arbitrary spin occupations and mass ratios, to 
second order in perturbation theory. Our main 
result is a closed analytical expression for the beyond-mean-field 
equation of state. To our knowledge, such an analytical expression 
has not been previously obtained for the mass-imbalanced case. 
%The full 
%derivation of this result, together with the global minimization procedure 
%used 
%to determine the optimal occupations ${C_\lambda}$, is provided in the 
%Supplemental Material.
Our analytical equation of state provides a unified framework to describe how mass imbalance modifies itinerant ferromagnetism in multicomponent gases. We find 
three robust features: (i) mass imbalance induces partial polarization already 
at vanishing coupling, (ii) heavier species are always favored energetically, 
and (iii) the interaction strength required to reach full spin polarization is 
significantly reduced, vanishing in the limit of extreme mass ratio.

%\section{Methodology}
We consider an heterogeneous SU(N) Fermi gas at zero temperature with repulsive 
short-range 
interactions characterized by a positive $s$-wave scattering length $a_0$. The 
populations of the spin components are written in terms of occupation factors 
$C_\lambda$, defined through $N_\lambda=C_\lambda N/\nu$, where $\nu=2S+1$ and 
$\sum_\lambda C_\lambda=\nu$. The Fermi momentum of the unpolarized gas is 
$k_F=\Big(6\pi^2 n/\nu \Big)^{1/3}$, which depends only on the density $n$ and 
spin degeneracy $\nu$, not on the masses. Mass 
imbalance 
enters through the ratios $r_\lambda=m_g/m_\lambda$, where $m_g$ is chosen as 
the heaviest mass in the system. The corresponding Fermi energy is 
$\varepsilon_F=\hbar^2 k_F^2/(2m_g)$.

Using perturbation theory to second order in the gas parameter $x=k_F a_0$, the 
energy per particle can be written as
\begin{equation}
\begin{aligned}
     \frac{E}{N}=\frac{3}{5}\varepsilon_F\bigg\{\frac{1}{\nu}\sum_{\lambda}r_{\lambda}C_{\lambda}^{5/3}\\
   +\frac{1}{\nu}\sum_{\lambda_1,\lambda_2}(1-\delta_{
\lambda_1,\lambda_2})\bigg[\frac{5}{9\pi}(r_{\lambda_1}+r_{\lambda_2})C_{\lambda_1}C_{\lambda_2}(k_Fa_0)\\+\frac{5}{4\pi^2}(r_{\lambda_1}+r_{\lambda_2})^2I_2(C_{\lambda_1}^{1/3},C_{\lambda_2}^{1/3},r_{\lambda_1},r_{\lambda_2})(k_Fa_0)^2\bigg]\bigg\} \ ,
\end{aligned}
\label{eq.forfinal}
\end{equation}
where $I_2\equiv 
I_2(C_{\lambda_1}^{1/3},C_{\lambda_2}^{1/3},r_{\lambda_1},r_{\lambda_2})$ is 
the second-order integral that encodes the beyond-mean-field corrections, 
originally derived for equal masses in Refs.~\cite{Pera1,Kanno1970}. To obtain a fully analytical expression for the energy, we integrate the function $I_2(k,l,r_1,r_2)$, whose expression is
\begin{equation}
\begin{aligned}
    I_2(k,l,r_1,r_2)=\frac{1}{16\pi^3}\int d\textbf{p}\, n_p\int d\textbf{p}'\,n_{p'}\\
    \times\int 2 \, d\textbf{q}\,d\textbf{q}'\frac{[n_q+n_{q'}-n_qn_{q'}]\delta(\textbf{q}+\textbf{q}'-\textbf{p}-\textbf{p}')}{r_1q^2+r_2{q'}^2-r_1p^2-r_2{p'}^2} \ .
\end{aligned}
\label{i2tot_mass}
\end{equation}

The analytical evaluation of $I_2$ (\ref{i2tot_mass}), for any 
Fermi momenta and arbitrary masses, is cumbersome and, to our knowledge, has not been reported previously. 
By defining  
$k=C_{\lambda_1}^{1/3}$, $l=C_{\lambda_2}^{1/3}$, and $r=r_1/r_2$, we obtain the 
closed expression~\cite{supplement}
\begin{widetext}
\begin{equation}
\begin{aligned}
    I_2(k,l,r_1,r_2)
    =\frac{1}{105r_2}\bigg\{\frac{2kl\Big[2l^5+rk(3k^4+22k^2l^2+l^4)-r^2l(3l^4+22l^2k^2+k^4)-2r^3k^5\Big]}{r(1+r)(1-r)}\\
    -\frac{4}{r^2}\bigg[l^7\ln{\bigg|\frac{l+k}{l}\bigg|}+r^3k^7\ln{\bigg|\frac{l+k}{k}\bigg|}-\frac{(l+rk)^5(l^2-5rkl+r^2k^2)}{(1+r)^2(1-r)^2}\ln{\bigg|\frac{l+k}{l+rk}\bigg|}\bigg]\\
    +\frac{(k-l)^4\Big[(4+3r)(4kl^2+l^3)+(3+4r)(4lk^2+k^3)\Big]}{(1+r)^2}\ln{\bigg|\frac{l-k}{l+k}\bigg|}
    -\frac{4r(35k^3l^4+14r^2k^5l^2-r^4k^7)}{(1+r)^2(1-r)^2}\ln{|r|}\bigg\} \ .
\end{aligned}
\label{i2final}
\end{equation}
\end{widetext}

This expression is finite in all physically relevant limits. In particular, it 
reproduces 
the known equal-mass result,  $r_1=r_2$~\cite{Pera1,Kanno1970}. The regular 
behavior of Eq. (\ref{i2final}) in limits such as $k=l$ or extreme mass 
imbalance provides a strong internal check of the calculation. For instance, 
in the case of the extreme mass imbalance limit $r_1\rightarrow0$, which is 
also $r\rightarrow0$, one obtains
\begin{equation}
\begin{aligned}
    I(k,l,0,r_2)=\frac{1}{105r_2}\bigg[6k^6l+44k^4l^3\\+2k^2l^5-8l^5(7k^2-l^2)\ln{\bigg|\frac{l+k}{l}\bigg|}\\
    +(k-l)^4(3k^3+12k^2l+16kl^2+4l^3)\ln{\bigg|\frac{l-k}{l+k}\bigg|}\bigg] \ .
\end{aligned}
\end{equation}
If, furthermore, we take $k=l$ in that limit, it simplifies to
\begin{equation}
    I(k,k,0,r_2)=\frac{2k^7}{35r_2}(9-8\ln{2}) \ .
\end{equation}
We omit the $r_2\rightarrow0$ limit because it follows straightforwardly by replacing $r_2$ with $r_1$ and swapping $k$ and $l$.

If Eq. (\ref{i2final}) is applied to the equal mass problem, that is $r_1=r_2$, 
which means $r=1$, we recover the known expression~\cite{Pera1,Kanno1970},
\begin{equation}
\begin{aligned}
    I(k,l,r_1,r_1)=\\ \frac{1}{420r_1}\bigg[2kl(k+l)(15k^4-19k^3l+52k^2l^2-19kl^3+15l^4)\\
    +7(k+l)(k-l)^4(k^2+3kl+l^2)\ln{\bigg|\frac{l-k}{l+k}\bigg|}\\
    -16k^7\ln{\bigg|\frac{l+k}{k}\bigg|}-16l^7\ln{\bigg|\frac{l+k}{l}\bigg|}
    \bigg] \ .
\end{aligned}
\end{equation}
If, still in the equal mass situation, we take  $k=l$ we reproduce the 
well-known result~\cite{Huang1957,Lee1957,Galitskii1958},
\begin{equation}
    I(k,k,r_1,r_1)=\frac{4k^7}{105r_1}(11-2\ln{2}) \ .
\end{equation}
Finally, for the case  $k=l$ and any mass ratio, we obtain
\begin{equation}
\begin{aligned}
    I(k,k,r_1,r_2)=\frac{4k^7}{105r_2}\bigg[\frac{1+14r+r^2}{r(1+r)}-\frac{1+r^3}{r^2}\ln{2}\\+\frac{(1+r)^3(1-5r+r^2)}{r^2(1-r)^2}\ln{\bigg|\frac{2}{1+r}\bigg|}\\
    -\frac{r(35+14r^2-r^4)}{(1+r)^2(1-r)^2}\ln{|r|}\bigg] \ .
\end{aligned}
\label{anyratio}
\end{equation}
Notice that this latter limit (\ref{anyratio}) corresponds to the case in which 
two spin species have the same populations, but differ in mass.

%\section{Results}

Before exploring the effect of interactions in the energy, we can inspect how 
mass imbalance already breaks the symmetry at very low $k_Fa_0$. Mass imbalance 
modifies the preferred configuration because the kinetic energy depends 
explicitly on the particle masses. Minimizing the non-interacting energy under 
the constraint $\sum_\lambda C_\lambda=\nu$ yields
\begin{equation}
    C_i=\frac{\nu\, r_i^{-3/2}}{\displaystyle\sum_{j=-s}^sr_j^{-3/2}}=
    \frac{\nu\, m_i^{3/2}}{\displaystyle\sum_{j=-s}^sm_j^{3/2}} \ ,
    \label{eq_C_mass}
\end{equation}
which shows that heavier species are already favored when $x \to 0$. 
Consequently, the reference 
configuration of the system is no longer the usual paramagnetic state, with 
equal populations, but an intrinsically polarized one determined by the mass 
hierarchy. We note that this minimization does not imply that particles change their mass; it only determines which distribution of the different species is energetically favored.

\begin{figure}[t]
    \centering
    \includegraphics[width=0.99\linewidth]{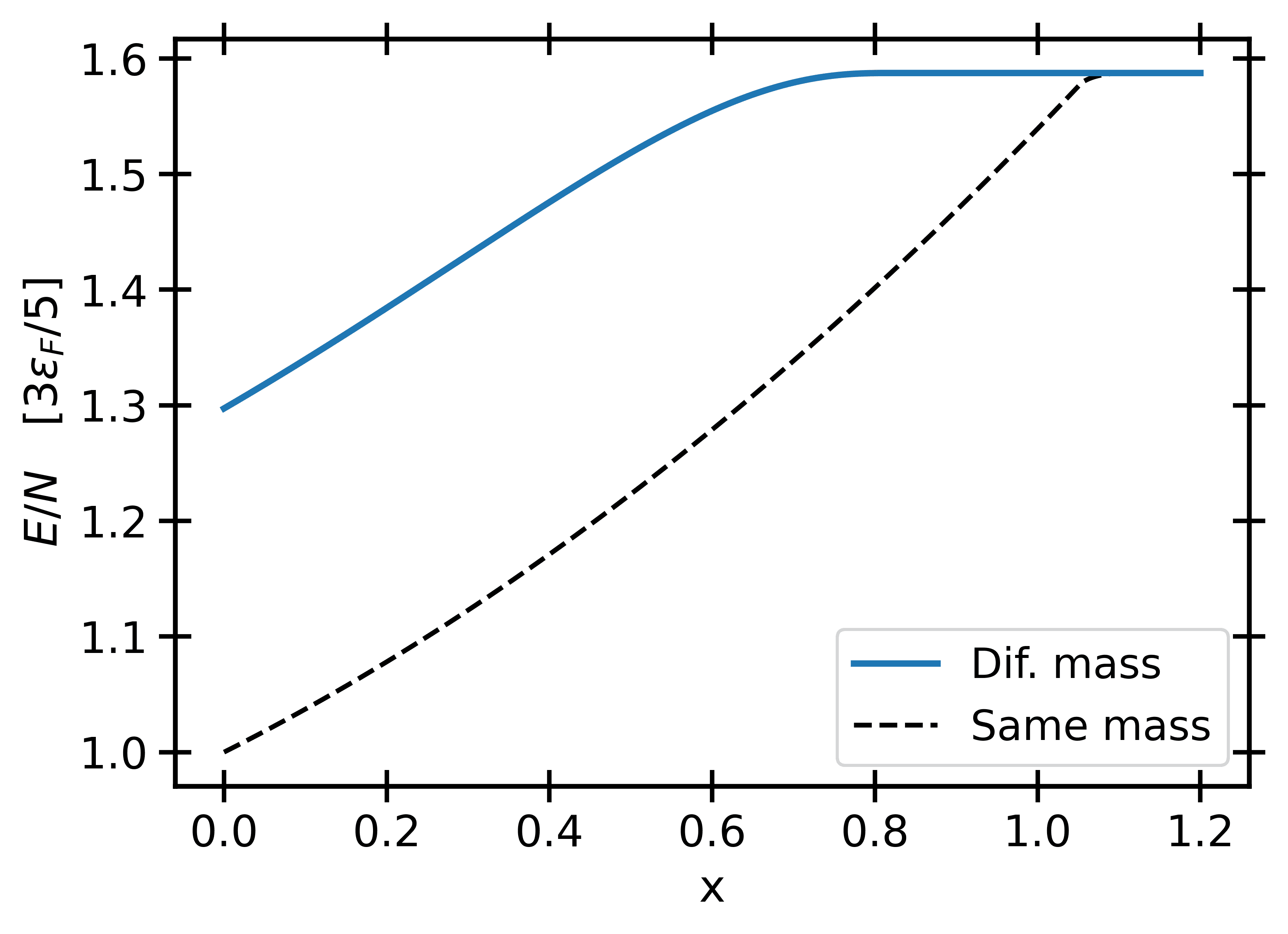}
    \caption{Energy for $S=1/2$ as a function of the gas parameter $x$. The 
black data represents the mass balanced configuration. In the mass imbalanced 
system, the spin component $C_{1/2}$ has a mass ratio of  $r_{1/2}=1$, while 
the other spin $C_{-1/2}$ has $r_{-1/2}=2$.}
    \label{fig_Epolm_0}
\end{figure}

To determine the magnetic behavior of the system, we minimize the second-order 
energy
globally with respect to the set ${C_\lambda}$ for fixed interaction strength 
$x$, spin degeneracy, and mass pattern. The analytical expression for $I_2$ 
(\ref{i2final}) makes this optimization efficient and numerically 
achievable~\cite{supplement}. 
We first analyze the simplest case $S=1/2$. The energetic effect of mass 
imbalance is illustrated in Fig.~\ref{fig_Epolm_0}, where the 
ground-state energy of a mass-balanced gas is compared with that of a mixture 
in which one component is twice as heavy as the other. Because the kinetic 
energy scales inversely with the particle mass, populating lighter species is 
energetically more costly. As a result, the mass-imbalanced system evolves 
toward the fully polarized configuration at a smaller interaction strength 
$x$ than the balanced gas.

\begin{figure}[t]
    \centering
    \includegraphics[width=0.99\linewidth]{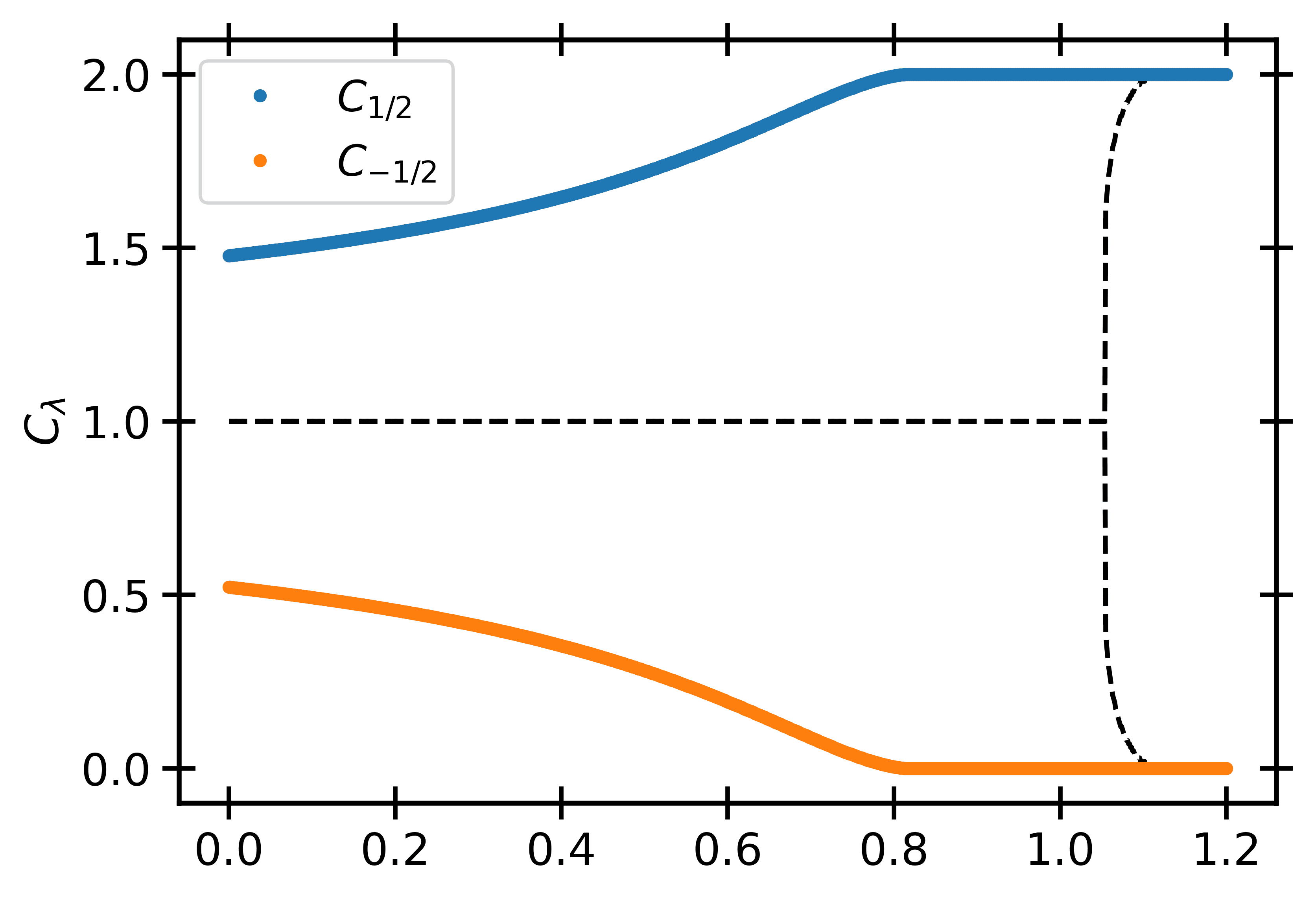}\\[-0.56cm]
    \includegraphics[width=0.99\linewidth]{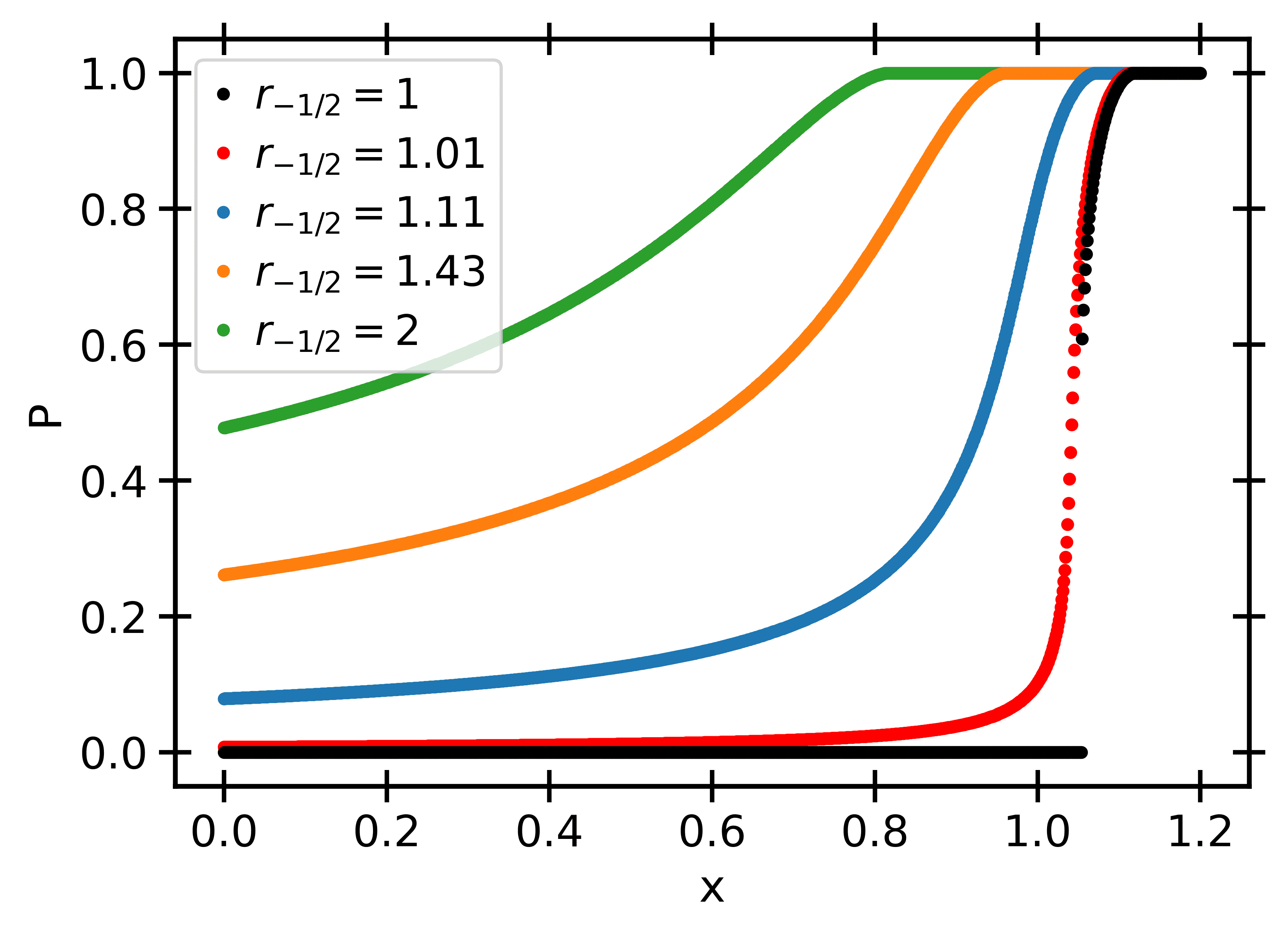}
    \caption{Occupation factors (top) and the corresponding polarization 
(bottom, green curve) for $S=1/2$ as a function of the gas parameter $x$. The black data 
represent the mass balanced configuration. The spin component $C_{1/2}$ has a 
mass ratio of $r_{1/2}=1$, whereas $C_{-1/2}$ has $r_{-1/2}=2$ in the top panel and $r_{-1/2}=1$, $1.01$, $1.11$, $1.43$, and $2$ in the bottom panel.}
    \label{fig_Cm_0}
\end{figure}

In order to understand better this behavior, we plot in Fig.~\ref{fig_Cm_0}
the occupation factors and the corresponding polarization (green curve). Even at 
weak coupling, the occupations differ from the equal-population configuration 
due to the mass asymmetry. As interactions increase, configurations with a larger occupation of the heavier species become energetically favored and the system evolves smoothly toward a fully polarized state. This does not imply a conversion between species with different masses; rather, it reflects a change in the energetically preferred occupation numbers of the different species. It is worth noticing that, in contrast with the 
discontinuous transition 
predicted for the mass-balanced gas 
within second-order perturbation theory, the mass-imbalanced system approaches 
full polarization continuously, from the already initially polarized 
configuration. The discontinuous transition only exists in the mass-balanced case.

\clearpage

\begin{figure}[t]
    \centering
    \includegraphics[width=1\linewidth]{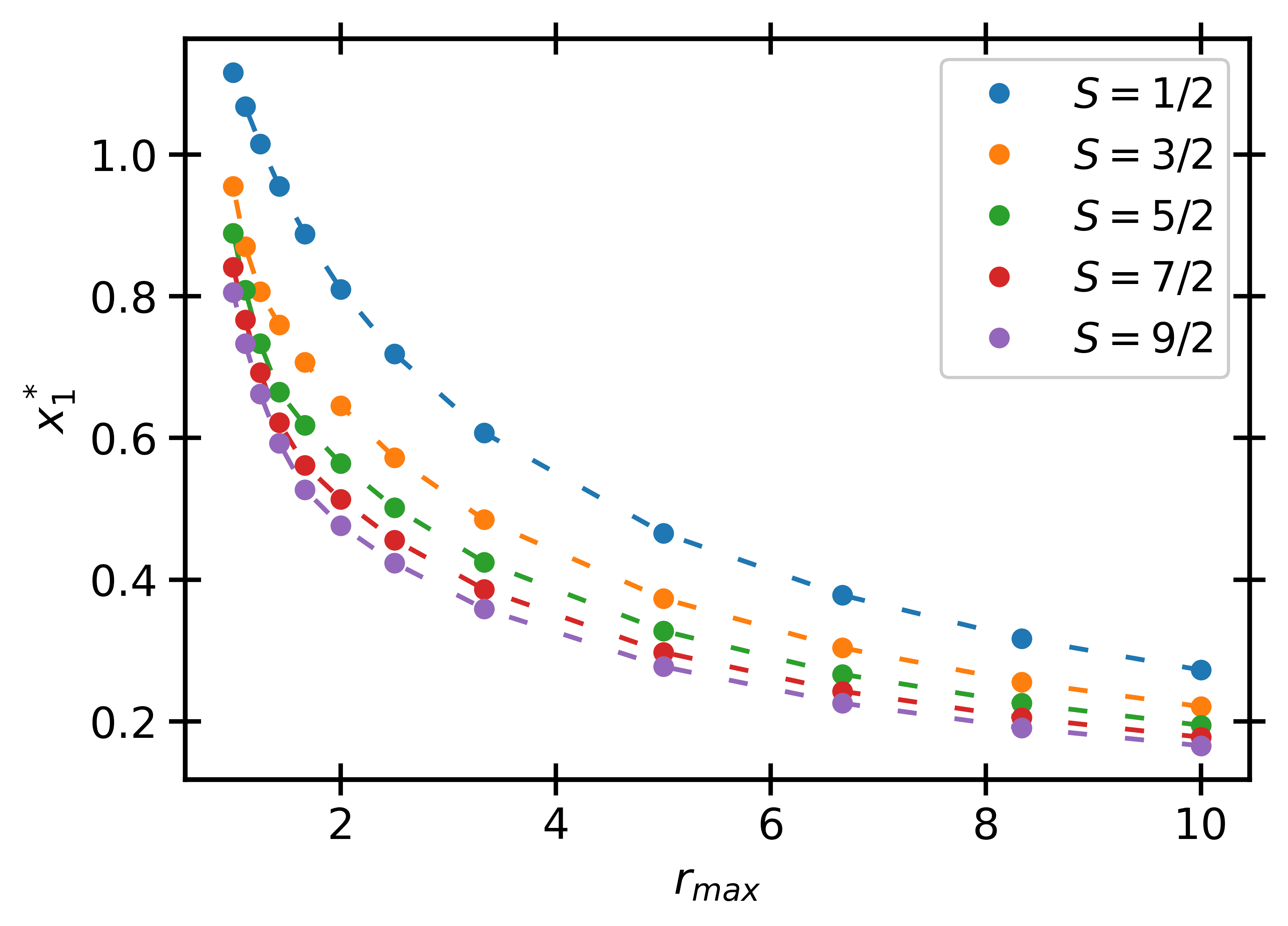}
    \caption{Gas parameter at which the system reaches $P=1$ as a function of the maximum mass ratio $r$ for $S=1/2$, $3/2$, $5/2$, $7/2$ and $9/2$. In all cases, one component is the heaviest one, with $r_s=1$, while the remaining components have $r_{\lambda\neq s}=r_{\rm max}$.}
    \label{fig_x1m_1}
\end{figure}

Unequal-mass mixtures shift ferromagnetism to lower gas parameters.
In Fig.~\ref{fig_x1m_1}, we show the gas parameter $x_1^*$ at which the system 
reaches the fully polarized state for several spin degeneracies. In all cases, 
we consider the same mass pattern: a single component has a different mass and 
is chosen to be the heaviest one, with $r_s=1$, while the remaining components 
have $r_{\lambda\neq s}=2$. As one can see, the critical value $x_1^*$ 
decreases monotonically as the mass ratio increases, showing that the promotion 
of ferromagnetism by mass imbalance is not restricted to $S=1/2$ but persists 
throughout the SU(N) Fermi systems considered here.

\begin{figure}[t]
    \centering
    \includegraphics[width=1\linewidth]{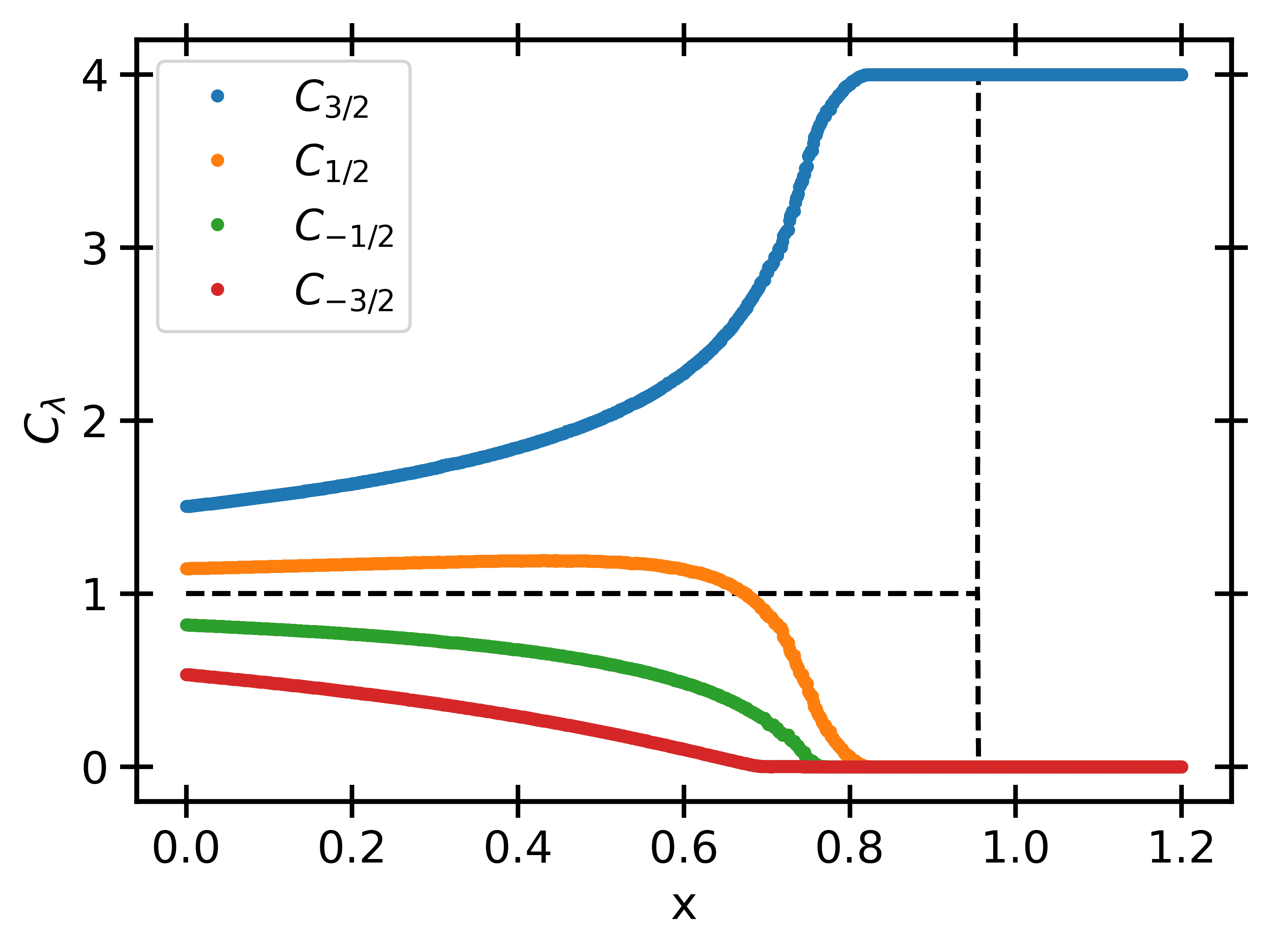}
    \caption{Occupation factors for $S=3/2$ as a function of the gas parameter 
$x$. The black dashed line represents the mass-balanced configurations. Each 
spin component $\lambda$ has a different mass: from the heavier $r_s=1$ to the 
lighter $r_{-s}=2$.}
    \label{fig_Cm_3}
\end{figure}

Focusing on large-spin mixtures, it is easy to understand that 
 one faces a much richer phenomenology because there are many more ways of 
assigning different masses, and therefore distinct polarization 
paths. In the following, we will discuss two scenarios: (i) all species have a 
different mass, and (ii) all species have the same mass, but one, which is 
lighter.
We first consider 
the situation where all spin components have different masses. 
In Fig.~\ref{fig_Cm_3}, we show an example for $S=3/2$, where the masses are 
graded regularly from $r_s=1$ to $r_{-s}=2$. In this configuration, the 
degeneracy is completely broken and each component follows a distinct trajectory 
as interactions increase. When $x$ increases, the concentrations of light 
species decrease progressively, following the value of their mass, and eventually the optimal configuration is occupied only by the heaviest species.

\begin{figure}[t]
    \centering
    \hspace*{0.26cm}\includegraphics[width=0.97\linewidth]{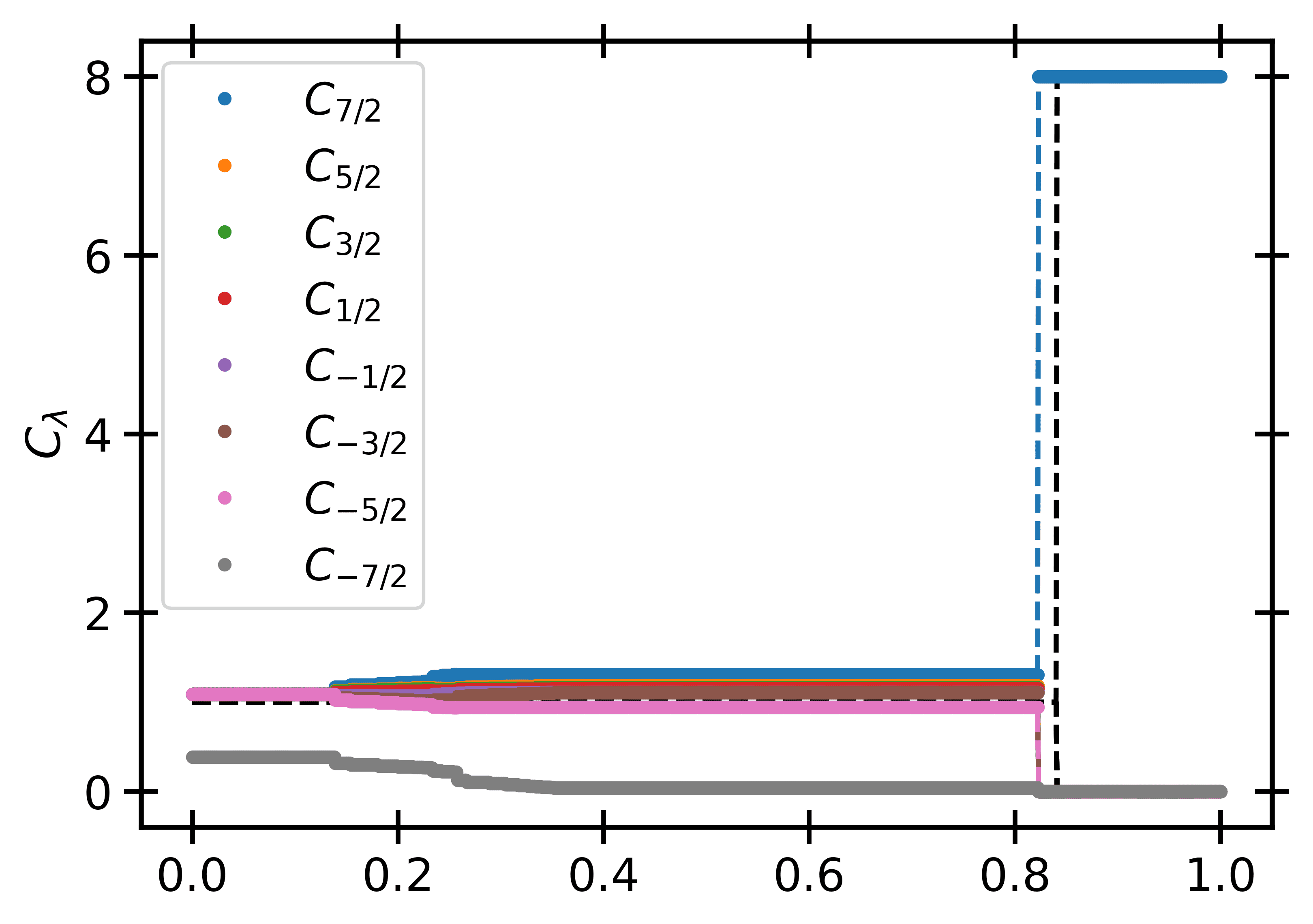}\\[-0.55cm]
    \includegraphics[width=1\linewidth]{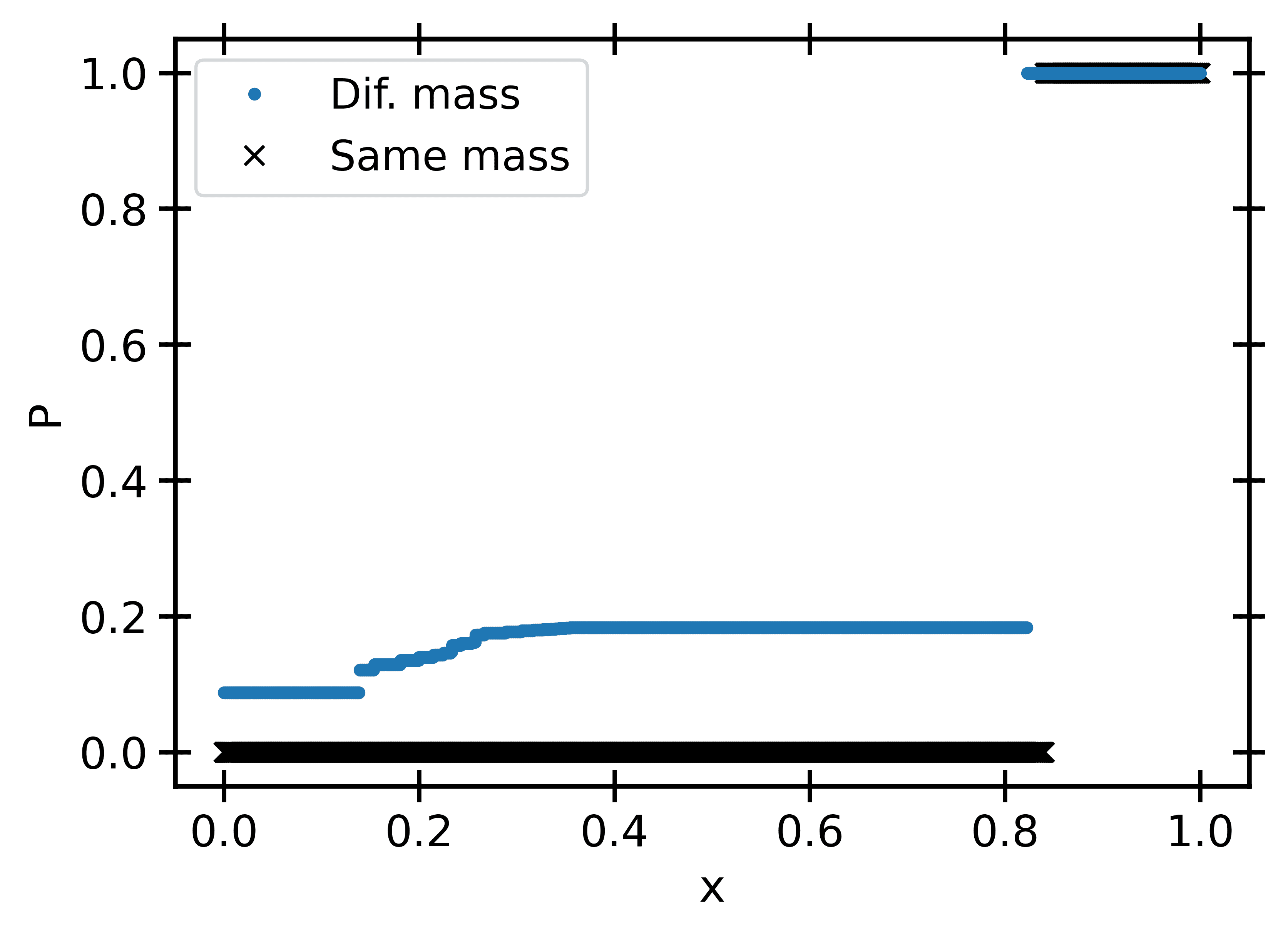}
    \caption{Occupation factors (top) and the corresponding polarization (below) 
for $S=7/2$ as a function of the gas parameter $x$. The black data represent 
the mass-balanced configuration. The spin component $C_{-s}$ has a different 
(light) mass, $r_{-s}=2$. The rest has $r_{\lambda\neq-s}=1$.}
    \label{fig_Cm_2}
\end{figure}

A qualitatively different scenario occurs when only one component is lighter 
than the rest. 
We have chosen it to be half of the mass of the others. In this case, the heavy 
species remain nearly degenerate and can sustain partially polarized plateaus 
before the final transition to full polarization. In Fig.~\ref{fig_Cm_2}, we 
illustrate this behavior for $S=7/2$ through the occupation factors and the 
resulting  polarization, where a stable partially polarized regime appears over 
a finite range of interaction strengths.

These results demonstrate that mass imbalance does not simply shift the 
ferromagnetic 
threshold but fundamentally modifies the route toward magnetic ordering. 
Depending on the mass hierarchy, the system may exhibit smooth polarization, 
weak discontinuities, or stable partially polarized regimes.

%\section{Conclusions}
Summarizing, we have derived the full analytical second-order energy of a 
repulsive mass-imbalanced 
SU(N) Fermi gas for arbitrary polarization and arbitrary mass ratio. The key 
result is a closed expression for the beyond-mean-field integral $I_2$, which 
had remained unavailable in analytical form for unequal masses.
This analytic equation of state shows that mass imbalance induces 
polarization already in the non-interacting limit, favors the occupation of the heavier components, 
and lowers the coupling needed to reach full ferromagnetism. Contrarily to 
the equal-mass mixture for $S=1/2$ at second order, the transition 
from the initial already asymmetric state toward the fully polarized one is 
continuous. For higher spin, different mass patterns generate qualitatively 
different magnetic paths, including smooth evolutions, discontinuous 
rearrangements, and stable partially polarized regimes. These findings identify 
mass engineering as a powerful strategy for controlling magnetism in ultracold 
Fermi gases. Already produced heterogeneous Fermi 
mixtures~\cite{Kohstall2012,Lippi2024, 
Ravensbergen2020}, and possibly future ones, can be exploited to observe the 
new phenomenology that our theory has proven.

\section{Acknowledgments}
We acknowledge financial support from Ministerio de Ciencia e Innovaci\'on
MCIN/AEI/10.13039/501100011033
(Spain) under Grant No. PID2020-113565GB-C21.

\bibliography{refs}% Produces the bibliography via BibTeX.

\end{document}